\documentclass[12pt,fleqn,a4paper]{article}
\pdfoutput=1
\usepackage{graphicx}
\usepackage{epsfig,cite}
\usepackage{amssymb}
\usepackage{amsmath}
\usepackage{color}
\usepackage{amstext,alltt,setspace}
\usepackage{amsbsy}
\usepackage{array}
\usepackage{booktabs, makecell}
\usepackage{hyperref}
\usepackage{slashed}
\usepackage{url}
\usepackage{geometry}
\providecommand{\pDa}[1]{{
\left(
\frac{1}{1-#1}
\right)_+
}}

\newcommand{\msbar}{$\overline{\rm { MS}}$}
\newcommand{\mmsbar}{{\overline{\rm {MS}}}}
\newcommand{\NS}{{\rm NS}}

\begin{document}
\begin{flushleft}
  \includegraphics[width=.3\linewidth,clip]{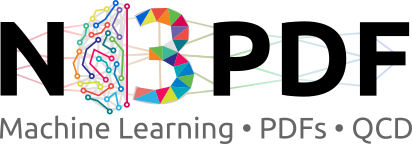}
\end{flushleft}
\vspace{-2.5cm}
\begin{flushright}
  \includegraphics[width=.15\textwidth]{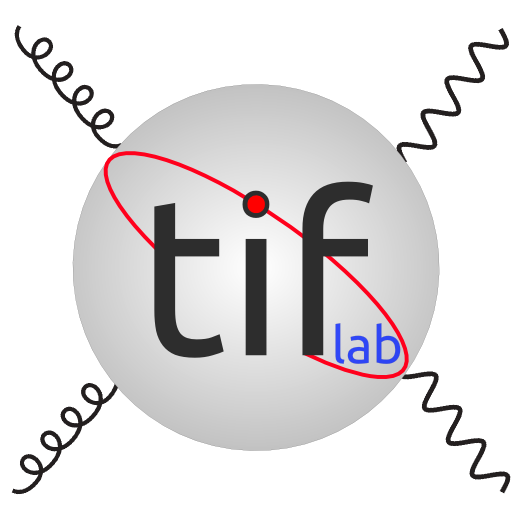}\\
  TIF-UNIMI-2023-21\\
\end{flushright}

\vspace*{.2cm}

\begin{center}
  {\Large \bf{On the positivity of \msbar{} parton distributions }}
\end{center}

\vspace*{.7cm}

\begin{center}
  Alessandro Candido$^1$, Stefano Forte$^1$, Tommaso Giani$^2$,  and Felix Hekhorn$^{1,3,4}$
  \vspace*{.2cm}

{  \noindent
      {\it
        ${}^1$Tif Lab, Dipartimento di Fisica, Universit\`a di Milano and\\ 
        INFN, Sezione di Milano,
        Via Celoria 16, I-20133 Milano, Italy\\[0.1cm]
    $^2$Department of Physics and Astronomy, Vrije Universiteit,
        NL-1081 HV Amsterdam and\\
Nikhef Theory Group, Science Park 105, 1098 XG Amsterdam, The Netherlands\\[0.1cm]
    ${}^{3}$University of Jyvaskyla, Department of Physics,\\
       P.O. Box 35, FI-40014 University of Jyvaskyla, Finland\\[0.1cm]
    ${}^4$Helsinki Institute of Physics, P.O. Box 64, FI-00014 University of Helsinki, Finland
}}

      \vspace*{3cm}

      {\bf Abstract}
\end{center}

{\noindent We revisit our argument that shows that  parton
  distribution Functions 
  (PDFs) in the \msbar{} scheme are non-negative in the
  perturbative region, with the dual goals of clarifying its
  theoretical underpinnings and elucidating its domain of
  validity. We specifically discuss recent results
  proving that PDFs can turn negative at sufficiently low
  scale, we  clarify quantitatively various aspects
  of our derivation of positivity in the perturbative region, and we
 provide  an estimate for the scale above which PDF positivity holds.
  }

\pagebreak

\section{PDF positivity}
\label{sec:intro}
The issue of positivity of parton distributions  raises interesting
questions both from a theoretical and a phenomenological point of
view. Leading-order (LO)  parton distributions are necessarily positive
because they are proportional to physically observable cross-sections,
while beyond LO positivity depends on the choice of
factorization scheme\footnote{Note that here
and henceforth we use the words ``positive'' and ``positivity'' to
mean more precisely ``non-negative''}. Hence, PDF positivity goes to
the heart of factorization; furthermore, PDF positivity in the
commonly used \msbar{} factorization scheme is phenomenologically
relevant for PDF determination.

In Ref.~\cite{Candido:2020yat} we have provided arguments for
perturbative positivity of PDFs in the \msbar{} factorization
scheme. Perturbative here means that these arguments apply in the
domain in which leading-twist perturbative factorization holds, and
moreover the perturbative expansion is well-behaved, meaning that
the size of perturbative corrections decreases with increasing
perturbative order. They are  
based on
starting with a scheme in which PDFs are positive, and then showing
that positivity is preserved when transforming to
\msbar{}, if the scheme change is perturbative, namely in a region in
which next-to-leading order (NLO) corrections are  smaller
than LO terms.  Specifically, as candidate starting positive schemes we
considered two possibilities:  a
scheme in which partonic 
cross-sections are positive beyond LO, and a physical
scheme in which PDFs are identified with physical observables. In the
former scheme, positivity holds provided the $d$-dimensional 
PDF  is positive before collinear subtraction. In the latter scheme
positivity is guaranteed at all scales by that of the physical observable.

These two arguments were argued in Ref.~\cite{Candido:2020yat} to be
equivalent. However, while positivity 
of the physical observable is guaranteed at all scales, it has been
recently shown 
in Ref.~\cite{Collins:2021vke} that positivity of the $d$-dimensional
collinear-unsubtracted (called ``bare'' PDF in
Ref.~\cite{Candido:2020yat})
cannot hold at all scales, and
must necessarily fail at a low-enough scale. 
This then raises the issue of the relation
between the two routes 
to positivity of Ref.~\cite{Candido:2020yat}, and more in general, the
question of whether the argument of Ref.~\cite{Collins:2021vke}
invalidates the positivity argument of Ref.~\cite{Candido:2020yat}, or
perhaps it restricts its domain of validity. This, in turn, raises
the question of the precise domain of validity of positivity: as
mentioned, in
Ref.~\cite{Candido:2020yat} it was shown to hold in the perturbative
region, but no precise assessment of this region was given.

It is the purpose of this note to address all these questions in
turn. The argument of Ref.~\cite{Candido:2020yat} is based on
two steps, first the construction or choice of a positive
factorization scheme, and then the transformation from this scheme to
\msbar{}, and consequently a general assessment requires revisiting
both steps. 
Specifically, 
in Section~\ref{sec:fact} we address the more conceptual issue of
understanding if and where positivity of the PDF holds, also in light
of the implications of
the argument of Ref.~\cite{Collins:2021vke} for the construction  of
a scheme with positive partonic cross-sections of
Ref.~\cite{Candido:2020yat}. 
In  Section~\ref{sec:pospos} we instead address the pragmatic issue of
determining whether PDFs in the \msbar{} scheme are positive or not,
by studying the transformation to  this scheme from a scheme in which
PDFs are surely positive, namely, a physical factorization scheme, and
assessing
phenomenologically the domain of applicability of positivity
conditions. The implications of our results, specifically to the
concrete issue of PDF determination, are discussed in Section~\ref{sec:conc}.

\section{Factorization: bare and renormalized PDFs}
\label{sec:fact}

The first route to proving  positivity in the \msbar{} scheme
considered in Ref.~\cite{Candido:2020yat} is to construct a
factorization scheme in which partonic cross-sections  are positive,
by a suitable 
modification of the  \msbar{} collinear subtraction.
This construction is rooted in  the  \msbar{}  factorization of
collinear singularities,  
most easily formulated for deep-inelastic
scattering in which case it can be derived from the Wilson expansion.

We now
summarize this factorization argument; in particular we make explicit
and general the dependence on the 
various scales, for which in Ref.~\cite{Candido:2020yat} implicit
choices were made. The factorization
consists of writing deep-inelastic scattering (DIS) structure
functions as
\begin{align} \label{eq:refaca}
 \frac{1}{x} F(x,Q^2,\epsilon)&=\sum_{j} e^2_j
 C_j\left(\alpha_s(\mu_r), \frac{Q^2}{\mu_r^2}\right) \otimes f_j(\mu_r^2)+O\left(\frac{\Lambda^2}{Q^2}\right) \\
&=\sum_{j} e^2_j C_j^R \left(\alpha_s(\mu_r), \frac{\mu^2_r}{\mu^2_f},\frac{Q^2}{\mu_f^2}\right)\otimes f_j^{R}(\mu_f^2)+O\left(\frac{\Lambda^2}{Q^2}\right),
\label{eq:refacb}
\end{align}
where the coefficient functions $C$ (or partonic cross-sections)
are the partonic structure functions computed on external on-shell partons,
with all parton masses set to zero, $f_i$ is the PDF for the $i$-th parton
and $\otimes$ denotes convolution, i.e.\ $[f\otimes
  g](x)=\int_x^1\frac{dy}{y}f(x/y)g(y)$, and for simplicity we are
considering the case of photon-induced DIS.
Equations~(\ref{eq:refaca}-\ref{eq:refacb}) are written in
$d=4-2\epsilon$ dimensions in order to regulate singularities, to be
discussed shortly, but in the final factorized formula
Eq.~\eqref{eq:refacb} the $\epsilon \to 0$ limit can be taken and the
coefficient functions and PDFs remain
finite.

We stress that, as mentioned  in the introduction, all arguments
presented in this paper hold in the domain in which leading-twist perturbative
factorization holds. So for DIS
Equations~(\ref{eq:refaca}-\ref{eq:refacb}) are just the
leading order of a Wilson expansion, and thus they only hold up to
 $O\left(\frac{\Lambda^2}{Q^2}\right)$ higher twist corrections, that
would involve multiparton operators. Hence, all equations written in
this section should be understood to hold up to $O\left(\frac{\Lambda^2}{Q^2}\right)$ terms,
even though we stop repeating this each time.

In Eq.~\eqref{eq:refaca} $\mu_r$ is a renormalization point
at which ultraviolet singularities (UV) are subtracted and the PDF
$f_j(\mu^2_r)$ is
defined by a suitable UV renormalization condition, while the coefficient functions $C_j$
are affected by collinear singularities. These coefficient functions
and PDF were referred to as ``bare'' in Ref.~\cite{Candido:2020yat} in order
to indicate that factorization of collinear singularities
has not been performed yet. However,
as stressed in Ref.~\cite{Collins:2021vke}, the composite operator
whose matrix
element  defines the PDF has already
undergone  UV renormalization, so $f_j(\mu^2_r)$ is
a UV-renormalized quantity: indeed, the dependence of $\mu_r$ arises
as a consequence of this UV renormalization.
In Eq.~\eqref{eq:refacb} instead, $\mu_f$ is a factorization scale and the
collinear singularities that arise in the massless coefficient functions have been
factored in the PDF $f_j^{R}$, such that the coefficient function and
PDF are separately finite. 

In order to understand better the singularity structure of the
factorization formulae, Eqs.~(\ref{eq:refaca}-\ref{eq:refacb}), we
discuss the scale dependence in somewhat greater detail.
If factorization is derived from the Wilson expansion, the coefficient
functions in Eq.~\eqref{eq:refaca} are the inverse Mellin transform of
Wilson coefficients. Because of their universality, these can be
evaluated by replacing the proton target with a target in which
operator matrix elements are known. If one specifically picks a massless, on-shell 
quark or gluon target, then the coefficient function simply coincides
with the structure function for a free on-shell quark or gluon in the massless case. As
stressed in 
Ref.~\cite{Collins:2021vke}, this way of defining the coefficient
function (or Wilson coefficient) entails  the choice of a specific 
renormalization condition for the PDF, namely
the condition 
\begin{equation} \label{eq:freef}
  f_{j/i}(z,\mu_r^2)=\delta_{ij}\delta(1-z)\,,
\end{equation}
where  $f_{j/i}(z,\mu_r^2)$ is the PDF for a $j$-parton in a $i$-parton
target, evaluated on-shell and with massless partons. 
This condition implies  that indeed the coefficient function
Eq.~\eqref{eq:refaca} is just
the structure function of a free on-shell massless quark,  as computed in
Ref.~\cite{Candido:2020yat}, and it corresponds to a specific choice
of renormalization scheme, referred to as BPHZ' in
Ref.~\cite{Collins:2021vke}. 
The renormalization condition, Eq.~\eqref{eq:freef}, makes the factorization,
Eqs.~\eqref{eq:refaca}-\eqref{eq:refacb}, 
on which Ref.~\cite{Candido:2020yat} is based on and denoted as track-B in 
Ref.~\cite{Collins:2021vke}, 
consistent with the treatment of factorization given 
in Refs.~\cite{Collins:1989gx,Collins:2011zzd}, denoted as track-A in Ref.~\cite{Collins:2021vke}.

Because the coefficient function $C_j$ in
Eq.~\eqref{eq:refaca} is
evaluated for massless, on-shell quarks, it is affected by collinear
singularities. On the other hand, if the structure function is
computed in a hadron target it must be finite. This then implies that
the PDF $f_j(\mu_r^2)$ in
Eq.~\eqref{eq:refaca}  must also be divergent in four dimensions  in
the BPHZ' scheme when evaluated for a hadron target so as to cancel the
divergence of the coefficient function. Clearly this is a
generic property whenever one chooses a target such that the structure
function is finite.

In order to understand the divergence of the PDF, and the way it determines its
positivity properties, thereby
making contact with the argument of Ref.~\cite{Collins:2021vke}, we
compute  the PDF
of a quark in a free off-shell quark target at NLO in the massless case. 
Combined with the
(universal) coefficient function, this amounts to the perturbative computation of
the structure function of an off-shell quark, which is finite because
of the off-shellness of the target. The computation is thus akin to
that of an operator matrix element in a free quark state, but now
with the off-shellness taking care of the collinear singularity.

This PDF can be thought of as the ``perturbative component'' of
the proton PDF, by interpreting the off-shell target as the internal
line of a diagram that contributes to the structure function in a
hadron target.  The  computation is closely related to 
that of Sec.~8 of Ref.~\cite{Collins:2021vke}, also presented in
Ref.~\cite{Aslan:2022zkz}, which is however performed in  scalar
Yukawa theory, while here we wish to illustrate 
the point raised in~\cite{Collins:2021vke} while remaining in the
context of QCD. 

We thus consider
\begin{align} \label{eq:refacc}
 \frac{1}{x} F^{M_q}(x,Q^2)= e^2_q C_q\left(\alpha_s(\mu_r), \frac{Q^2}{\mu_r^2}\right) \otimes f^{M}_{q/q}\left(\mu_r^2\right)
 +\mathcal{O}(\alpha_s^2)
\end{align}
where $p^2=M^2$ is the target quark virtuality and
$f^{M}_{q/q}$ is the PDF for a quark in a quark target with
electric charge $e_q$, and henceforth all quantities are
determined up to NLO, i.e.\ neglecting terms that lead to
$\mathcal{O}(\alpha_s^2)$ contributions to the structure function.
The notation $F^{M_q}$ indicates that this is the
structure function of a free off-shell quark. Note that up to $O(\alpha_s^2)$ only diagonal
contributions, i.e.\ such as the struck quark is the same flavor as the
target quark, are present, but starting at $O(\alpha_s^2)$ the PDF
$f^{M}_{q_i/q_j}$ for
any quark of flavor $i$ in a target $j$-quark is nonvanishing so the
structure function has again the form of Eq.~\eqref{eq:refacc} with a
sum over quark flavors $j$.

The coefficient
function $C_q$ was given in
Ref.~\cite{Candido:2020yat}, Eq.~(2.24): it is
universal, hence the same regardless of whether the target is a
quark or indeed a hadron. Because the target 
is off-shell the PDF $f^{M}_{q/q}$ can only satisfy the condition
Eq.~\eqref{eq:freef} 
at LO: indeed, as already mentioned, beyond LO  $f^{M}_{q/q}$ must
necessarily contain a collinear singular contributions that cancels
against that of the coefficient function, leading to a finite
structure function. In order to compute it at NLO, and check the cancellation of
the collinear singularity,  we start with
a ``fully bare'' PDF $f^{M,\,0}_{q/q}$, and perform UV
renormalization in the BPHZ' scheme. This then gives the PDF $f^{M}_{q/q}$ of
Eq.~\eqref{eq:refacc}, corresponding to  $f_j(\mu_r^2)$ of
Eq.~\eqref{eq:refaca}: UV renormalized, but not yet collinear
factorized, and called ``bare'' in Ref.~\cite{Candido:2020yat}. From
this PDF the ``fully
renormalized'' PDF  $f^{M,\,R}_{q/q}$, to be defined precisely below, and
corresponding to $f_j^{R}(\mu_f^2)$ of Eq.~\eqref{eq:refacb}, 
is derived through  collinear factorization. We
henceforth stick to the terminology ``fully bare'' and ``fully
renormalized'' in order to refer to  $f^{M,\,0}_{q/q}$ and 
$f^{M,\,R}_{q/q}$ respectively. The singularity structure that we
expect, and that we verify explicitly, is the following: the
fully bare PDF  $f^{M,\,0}_{q/q}$, is UV divergent and IR finite; 
after UV renormalization, the PDF  $f^{M}_{q/q}$ is UV finite but
collinear divergent; the fully renormalized PDF $f^{M,\,R}_{q/q}$ is
finite. The real emission contribution to the
coefficient function $C_q$ is of course UV finite at NLO, it is IR finite
after addition of the virtual correction but it remains collinear
singular; its collinear singularity cancels against that of the PDF $f^{M}_{q/q}$
in Eq.~\eqref{eq:refacc}.

The NLO correction to the PDF $f^{M}_{q/q}$
involves~\cite{Collins:2011zzd} an integration 
over the momentum of the gluon radiated from the quark line, in which
the longitudinal and transverse momentum integrations are treated
asymmetrically, because the off-shell quark whose PDFs is being
determined  has a well-defined value of the longitudinal
momentum, which corresponds to that specified by the renormalization
condition Eq.~\eqref{eq:freef} that enforces it to all orders.
The  integral over the  transverse momentum
${\bf k}_\text{T}$ of the emitted quark then leads to the result
\begin{align}
  \label{eq:bare1}
  f^{M,\,0}_{q/q}\left(x,\mu^2, \epsilon\right)= \delta\left(1-x\right) + \frac{\alpha_s\left(\mu\right)}{2\pi}C_F \left\{
    \left[\Gamma\left(\epsilon\right)\left(\frac{\Delta^2\left(x\right)}{4\pi\mu^2}\right)^{-\epsilon}
    p_{qq}\left(x\right)\right]_+ +f_f(x)\right\}.
\end{align}
which holds in $d$ dimensions and is UV divergent as  $\epsilon\to0$
(see Appendix~\ref{sec:app1} for the details of the calculation).
In Eq.~\eqref{eq:bare1} $\mu$ is the standard dimensional
regularization scale;
$p_{qq}(x)$ is implicitly defined in terms of the quark-quark
splitting function $P_{qq}(x)$ as
$  P_{qq}(x)= C_F\left[p_{qq}(x)\right]_+$;  
$\Delta^2\left(x\right)$ is given by
\begin{align}
  \label{eq:mass_param}
  &\Delta^2\left(x\right) = -x\left(1-x\right)M^2\,;
\end{align}
and $f_f(x)$ is a UV finite and thus scale-independent contribution,
whose explicit form is given  in Appendix~\ref{sec:app1}, Eq.~\eqref{eq:finite}.

The UV divergence of the fully bare PDF, Eq.~(\ref{eq:bare1}), is a
consequence of the fact that the PDF is defined  as the
matrix element of a composite operator~\cite{Collins:1984xc}. The
divergence can be removed by a subtraction
\begin{equation}\label{eq:uvsubt}
 f^{M}_{q/q}\left(x,\mu_r^2, \epsilon\right)=
 f^{M,\,0}_{q/q}\left(x,\mu_r^2,
 \epsilon\right)-\delta^{\rm UV}\left(x,\mu_r^2,
 \epsilon\right),
\end{equation}
thereby leading to the UV renormalized, but still collinear divergent PDF
$f^{M}_{q/q}$.
The UV counterterm in Eq.~\eqref{eq:uvsubt} is fixed by the
renormalization condition 
Eq.~\eqref{eq:freef}, i.e.\ by considering Eq.~\eqref{eq:bare1} in the on-shell, massless case
and subtracting everything but the
$\delta(1-x)$ contribution to its r.h.s., with
$\mu=\mu_r$.
The explicit expression of $\delta^{\rm UV}$ is found performing the 
computation leading to Eq.~\eqref{eq:bare1}
in the on-shell, massless case,
i.e.\ starting from the expression of the bare PDF in terms of 
transverse momentum integrals (see  Appendix~\ref{sec:app1},
Eq.~\eqref{eq:bare}) and setting $M^2=0$. This leads to
\begin{align}
\label{eq:uvct}
  \delta^{\rm UV}\left(x,\mu_r^2,
  \epsilon\right)= \frac{\alpha_s\left(\mu_r\right)}{2\pi} &
  P_{qq}\left(x\right) \, 4\pi \mu_r^{2\epsilon}
  \int\frac{d^{2-2\epsilon} {\bf k}_\text{T}}{\left(2\pi\right)^{2-2\epsilon}} \frac{1}{k_\text{T}^2 }\,.
\end{align}

It is useful to separate the UV and IR regions of integration over
${\bf{k}}_\text{T}$ in Eq.~\eqref{eq:uvct}; this can be done by introducing an
auxiliary mass parameter on which nothing depends, and which we may
well take equal to the virtuality $M$, through the identity
\begin{align}   \label{eq:split_UV_and_IR}
  4\pi\mu_r^{2\epsilon} \int \frac{d^{2-2\epsilon} \bf{k}_\text{T}}{\left(2\pi\right)^{2-2\epsilon}}
  \frac{1}{k_\text{T}^2}  &= 
  4\pi\mu_r^{2\epsilon}  \int \frac{d^{2-2\epsilon} \bf{k}_\text{T}}{\left(2\pi\right)^{2-2\epsilon}} 
  \frac{1}{k_\text{T}^2 + M^2}
  + 4\pi\mu_r^{2\epsilon}  \int \frac{d^{2-2\epsilon} \bf{k}_\text{T}}{\left(2\pi\right)^{2-2\epsilon}} 
  \frac{M^2}{k_\text{T}^2\left(k_\text{T}^2+M^2\right)}  \nonumber \\
  &= \left(\frac{M^2}{4\pi \mu_r^2}\right)^{-\epsilon}\Gamma\left(\epsilon\right) -
  \left(\frac{M^2}{4\pi \mu_r^2}\right)^{-\epsilon}\Gamma\left(\epsilon\right).
\end{align}
Note that the first integral (UV divergent) converges in the
$\epsilon>0$ region, while the second one (IR divergent) converges
when  $\epsilon<0$. 

We thus get that the PDF at scale $\mu_r$ is given by
\begin{align}
  \label{eq:BPHZ'}
  f^{M}_{q/q}&\left(x,\mu_r^2, \epsilon\right)=
  \delta\left(1-x\right) + \frac{\alpha_s\left(\mu_r\right)}{2\pi}C_F 
    \left\{\left[\Gamma\left(\epsilon\right)
    \left(\frac{\Delta^2\left(x\right)}{4\pi\mu_r^2}\right)^{-\epsilon}
    p_{qq}\left(x\right)\right]_+
    +f_f(x)\right\} \nonumber \\
    & = \delta\left(1-x\right) + \frac{\alpha_s\left(\mu_r\right)}{2\pi}
    P_{qq}\left(x\right)\left(\frac{1}{\epsilon}-\gamma_E+\ln
    4\pi \right) + C_F\left\{\left[\ln\frac{\mu_r^2}{|\Delta^2\left(x\right)|}
    p_{qq}\left(x\right)\right]_+ +\bar f_f(x)\right\},
\end{align}
where $\bar f_f(x)=f_f(x)+i\arg(-M^2)$ now includes an imaginary
contribution when $M^2>0$ so the off-shell quark can decay into a
quark-gluon pair.
Note the different nature of the poles appearing in
Eqs.~\eqref{eq:bare1} and \eqref{eq:BPHZ'}: the UV divergence in Eq.~\eqref{eq:bare1} 
has been canceled by the UV divergent integral of
Eq.~\eqref{eq:split_UV_and_IR}, but the PDF has now acquired a
collinear divergence, which is the pole appearing in Eq.~\eqref{eq:BPHZ'}. 

We can finally determine the fully renormalized PDF. In
Ref.~\cite{Candido:2020yat} the collinear counterterm that removes the 
singularity from both the coefficient function and PDF was determined
by performing an \msbar{} subtraction of the coefficient
function. Here we can easily determine it from the expression of the
PDF, Eq.~\eqref{eq:BPHZ'}, by writing the fully renormalized PDF as
\begin{align}
  \label{eq:frpdf}
  f^{M,\,R}_{q/q}\left(x,\mu_f^2\right)=\lim_{\epsilon\to0}\left[f^{M}_{q/q}\left(x,\mu_r^2,
    \epsilon\right)+\delta^{\mmsbar}_q\left(x,\frac{\mu_f^2}{\mu^2_r},\epsilon\right)\right],
  \end{align}
where we have adopted the same sign convention for the counterterm as
in Ref.~\cite{Candido:2020yat}. It immediately follows that the
\msbar{} counterterm is
  \begin{align}
  \label{eq:collct}
  \delta^{\mmsbar}_q\left(x,\frac{\mu_f^2}{\mu^2_r},\epsilon\right)= \frac{\alpha_s\left(\mu_r\right)}{2\pi} 
  \left(\frac{\mu_f^2}{4\pi
  \mu^2_r}\right)^{-\epsilon} \left(-\frac 1
  {\epsilon}+\gamma_E\right) P_{qq}(x)\,,
\end{align}
and the fully renormalized PDF is 
 \begin{align}
  \label{eq:finq}
  f^{M,\,R}_{q/q}\left(x,\mu_f^2\right) 
  &=  \delta\left(1-x\right) 
  + \frac{\alpha_s\left(\mu_r\right)}{2\pi} C_F\left\{
  \left[\ln\frac{\mu_f^2}{|\Delta^2\left(x\right)|}p_{qq}\left(x\right)\right]_+
    +\bar f_f(x)\right\}.
\end{align}
Using the expression of the NLO coefficient function from
Ref.~\cite{Candido:2020yat} it is easy to check that the counterterm
also removes the collinear divergence from the coefficient function. 
The
renormalized coefficient function is 
\begin{align}\label{eq:crenorm}
  C^R_{q}&\left(x,\frac{Q^2}{\mu_f^2}\right)=\lim_{\epsilon\to0} 
  C_q\left(x,\left(\frac{Q^2}{\mu_r^2}\right) ,\epsilon\right)
  -\delta^{\mmsbar}_q\left(x,\frac{\mu_f^2}{\mu^2_r},\epsilon\right) \nonumber\\
  &= \delta(1-x)+ \frac{\alpha_s\left(\mu_r\right)}{2\pi}\left\{ 
  P_{qq}(x)\ln \frac{Q^2}{\mu^2_f}+ \left[C_F p_{qq}(x)
  \ln\left(\frac{1-x}{x}\right) \right]_+ + D(x)\right\},
\end{align}
with
\begin{align}
  \label{eq:ddef}
  D(x)= C_F\left[ -\frac {3}{2}\pDa{x} + 3
  + 2 x - 4 \delta(1-x)\right]\,.
\end{align}
The expression of the renormalized coefficient function of
Ref.~\cite{Candido:2020yat} coincides with  Eq.~\eqref{eq:crenorm}
with the choice $\mu^2_f=\mu_r^2=Q^2$.

The dependence of the PDF, Eq.~\eqref{eq:BPHZ'},
on  the  scale $\mu_r$, and consequently of the fully renormalized PDF,
Eq.~\eqref{eq:finq}, on  the  scale $\mu_f$, is driven by the splitting
function $P_{qq}$ as it should: the collinear singularity in the
coefficient function exactly matches the UV anomalous dimension
of the operator matrix element, i.e.\ of the PDF, in such a way that
the singularities in the coefficient function and PDF in
Eq.~\eqref{eq:refaca} cancel each other. Indeed, up to the 
finite terms, we could have
obtained Eq.~\eqref{eq:BPHZ'} 
by simply noting that (a) the dependence
of the PDF $f^{M}_{q/q}\left(x,\mu_r^2, \epsilon\right)$ on the scale
$\mu_r$ must satisfy the QCD evolution equation, i.e.\ a
Callan-Symanzik equation with anomalous dimension $P_{qq}$ 
(more precisely, its Mellin transform); 
(b) the dependence on $\mu_r$ must
cancel between the coefficient function and the PDF; 
(c) the dependence on scale of the coefficient function is due to its collinear singularity; 
(d) the PDF depends (on top of the
renormalization
scale) on a scale $\Delta^2$ that cuts off the collinear singularity.
The computation that we presented amounts to an explicit check
of all this. 

We now turn to the positivity properties of the PDF. As a preliminary
observation we note that the PDF of a quark is actually a
distribution, so we only discuss its positivity properties in the
$x<1$ region. Also, if $M^2>0$ the PDF develops an absorptive part, in
which case when discussing positivity we refer to the positivity of
the real part of the PDF.
Also,  we recall that our whole argument starts
with the leading-twist factorization
Eqs.~(\ref{eq:refaca}-\ref{eq:refacb}), hence at low enough scale 
any conclusion is  not only subject to large higher order perturbative
corrections, but also to sizable  higher-twist corrections.

All this said, we note that before collinear subtraction
the PDF $f^{M}_{q/q}$, Eq.~\eqref{eq:BPHZ'}, contains a singular
contribution proportional to
$\frac{P_{qq}\left(x\right)}{\epsilon}$, and  associate logarithmic
contribution proportional to
$P_{qq}\ln\frac{\mu_r^2}{|\Delta^2\left(x\right)|}$. But for
$x < 1$, $P_{qq}(x) > 0$, so the logarithmic term is negative at low
enough scale $\mu_r^2< |\Delta^2|$, and the PDF becomes negative at low
enough $\mu_r^2$  for any fixed value of $\epsilon<0$, in the
region of convergence of the transverse momentum integral. Correspondingly for
any fixed value of  $\mu_r^2$ the PDF becomes negative for
sufficiently small $\epsilon$. The interplay of $\epsilon$ and
$\mu_r^2$ in the positivity condition is of course due to the fact
that  in dimensional regularization with \msbar{}
subtraction $\mu=\mu_r$ plays the role of a cutoff. The fact that the PDF
$f^{M}_{q/q}$ is negative at low scale is the main observation of
Ref.~\cite{Collins:2021vke}. 

Turning now to the fully renormalized PDF
$f^{M,\,R}_{q/q}\left(x,\mu_f^2\right)$, Eq.~\eqref{eq:finq}, it is
clear that, since the logarithmic contribution is the same as before collinear
subtraction,  but with
now $\mu_r$ replaced by $\mu_f$, by the same argument  this
contribution will be dominant and the PDF will always be positive for
large enough $\mu_f$. Conversely,
$f^{M,\,R}_{q/q}\left(x,\mu_f^2\right)$ will always be negative for
small enough $\mu_f$, though of course at low scale both the NLO
perturbative approximation, and in fact the whole leading-twist
approach on which the computation is based fail. The transition
between these two regions depends on the finite terms, as also
emphasized in Sec.~8 of Ref.~\cite{Collins:2021vke} (see in particular Fig.~2),
where this result is discussed in the case of a Yukawa theory. 

Hence, positivity of the fully
renormalized PDF will surely hold only at high enough scale, such that $\mu_f$ is
an UV scale,  as it should be.
In a hadronic target the PDF cannot be computed explicitly, however
the same calculation leading to Eq.~\eqref{eq:finq} can be used to obtain 
the perturbative dependence of the PDF on the factorization scale. It
is the clear that the same arguments apply, but now with $\Delta^2\left(x\right)$
replaced by a characteristic scale $\Lambda^2_h$ of the hadronic target.
Namely, the renormalized quark PDF in a hadron target $f^{R}_q$ contains a logarithmic
contribution of the form of Eq.~\eqref{eq:finq}, i.e.\ it is given by
\begin{equation}\label{eq:hadpdf}
  f^{R}_q\left(x,\mu_f^2\right) =\delta(1-x)
+ \frac{\alpha_s\left(\mu_r\right)}{2\pi} 
  \ln\frac{\mu_f^2}{\Lambda^2_h}P_{qq}+ f_h(x)
\end{equation}
where  $f_h(x)$ is a scale-independent contribution determined by
non-perturbative physics, and the scale dependence is fully determined
by the QCD evolution equation.

We can now examine the positivity argument of
Ref.~\cite{Candido:2020yat} within the context of this model computation.
We start with the observation that 
combining Eqs.~\eqref{eq:crenorm} and~\eqref{eq:finq},
the structure function  $F^{M_q}(x,Q^2)$, Eq.~\eqref{eq:refacc}, when
$\epsilon=0$ and  $x<1$, is given by
\begin{align}
  \frac{1}{x}F^{M_q}(x,Q^2)&=
  e^2_q\frac{\alpha_s\left(\mu_r\right)}{2\pi}\left[
  P_{qq}(x)\ln\frac{Q^2(1-x)/x}{|\Delta^2\left(x\right)|} +  D(x)+C_F \bar f_f(x)\right]\\
  &=e^2_q\frac{\alpha_s\left(\mu_r\right)}{2\pi}C_F
  \left[\left((1+x^2)\ln\frac{Q^2}{|M^2|x^2}-\frac{3}{2}\right)\frac{1}{1-x}+
    3x + 1\right].
  \label{eq:finmfq}
\end{align}
The structure function is positive for all $x$ provided $Q^2$ is large
enough (the exact condition is $\frac{Q^2}{|M^2|}>
\exp\frac{3}{4}\approx 2$). This follows from the fact that
in this simple model the argument of the log is
$\frac{Q^2(1-x)/x}{|\Delta^2\left(x\right)|} =\frac{Q^2}{|M|^2 x^2}$.

The argument of Ref.~\cite{Candido:2020yat} then consists of noting
that in standard \msbar{} factorization, in which the coefficient
function is given by Eq.~\eqref{eq:crenorm} with $\mu^2_f=\kappa Q^2$
(or, more in general, $\mu^2_f$ is taken to be $x$-independent), the
coefficient function acquires a spurious logarithmic dependence on
$1-x$ because subtraction of the collinear
singularity  is performed at a fixed $x$-independent scale $Q^2$, rather than at the
physical scale of the process.
This is obvious in the model
computation: in \msbar{}
factorization the  logarithm is split according to
\begin{equation}
  \label{eq:splitlog}
  \ln\frac{Q^2(1-x)/x}{|\Delta^2\left(x\right)|}=\ln\frac{1-x}{x}+\ln\frac{Q^2}{|M^2| x (1-x)}\,,
\end{equation}
the first log is included in the coefficient function,
and the second in the PDF.
This amounts to  splitting a  positive log into the sum of a
negative contribution, included  in the coefficient function, and a
compensating contribution  included in the PDF. Consequently, the
NLO contribution to the coefficient function
may  turn negative. Note that in a perturbative approach, the
logarithm would be split
in an additive way, i.e. as $1+\alpha_s(\ln A+ \ln B)=(1+\alpha_s\ln
A)(1+\alpha_s \ln B)+ O(\alpha_s^2)$. If $\ln A$ is negative the
compensating $\ln B$ term is 
positive, so if the second factor is included in the PDF this
remains positive. However, if the negative logarithmic contribution
is large enough so $1+\alpha_s\ln
A$ turns negative, the perturbative approximation is not justified and  the split must be
treated multiplicatively, i.e. $1+\alpha_s(\ln A+ \ln B)=(1+\alpha_s\ln
A)\frac{1+\alpha_s(\ln A+ \ln B)}{1+\alpha_s\ln
A}$. The second factor is now negative and if included in the PDF the
latter
turns negative.
This happens both as a matter of principle if the PDF is computed
non-perturbatively (e.g. on the lattice), and
also in practice, if the PDF is extracted from the data, by
comparing the \msbar{} prediction with a negative coefficient function to 
the measured positive structure function.

In  Ref.~\cite{Candido:2020yat} it was consequently
suggested to perform factorization through   subtraction 
at the physical scale, in such a way that the coefficient
function remains positive. In such a scheme, the logarithmic
contributions that are associated to collinear divergences are split
between coefficient function and PDF in a way that preserves the
positivity of both. The  extra caveat, and main point of
Ref.~\cite{Collins:2021vke}, is that this ensures positivity of the
PDF  only provided
the scale is high enough, because otherwise the $d$-dimensional PDF
and the four-dimensional fully renormalized PDF are not positive to
begin with, so positivity is violated however one splits the log. 

This is  manifest in the model
computation: for $Q^2<|M^2|$ the logarithmic contribution is negative,
for small enough $Q^2$ it will overwhelm any positive finite term, and
the structure function
Eq.~\eqref{eq:finmfq} may turn negative. Hence even in
factorization
schemes such as those considered in Ref.~\cite{Candido:2020yat}
positivity of the PDF (and of the perturbative, leading-twist
structure function) only holds at high enough scale: physically, at
scales which are significantly larger than the 
hadronic 
target mass scale $\Lambda_h^2$ that regulates the collinear
singularity.

In summary, the import of the argument of
Ref.~\cite{Collins:2021vke} is that the construction of a scheme in
which coefficient functions are positive of
Ref.~\cite{Candido:2020yat} only leads to positive PDFs and positive
physical cross-sections at high enough scale. The value of this scale
is determined by non-perturbative physics, and this hampers a purely
perturbative determination of the range of validity of PDF
positivity. We address the
issue of the range of validity of positivity in the last section.

\section{From a physical scheme to \msbar{}}
\label{sec:pospos}

The issue of PDF positivity in the \msbar{} scheme can be cast in
purely perturbative terms by expressing the PDFs in terms of physical
observables, through a suitable choice of factorization scheme. 
This effectively  amounts to inverting Eq.~(\ref{eq:refacb}) at
leading twist.

In order to understand this schematically, we start by writing the factorized
expression Eq.~(\ref{eq:refacb})
for the nonsinglet structure function  $F_2^{\NS}$ in the  \msbar{}
scheme with $\mu_r^2=Q^2$:
\begin{align}\label{eq:disfact0}
 \frac{1}{x} F_2^{\NS}(x,Q^2)= \langle e_i^2\rangle C^{\rm NS,\,\mmsbar}\left(\alpha_s(Q^2), \frac{Q^2}{\mu_f^2}\right) \otimes 
 f^{\NS,\,\mmsbar}(\mu_F^2) +O\left(\frac{\Lambda^2}{Q^2}\right),
 \end{align}
where $f^{\NS}$ is a difference of two quark or antiquark PDFs, 
$\langle
e^2_i\rangle=\frac{1}{2} \left(e^2_i+e^2_j\right)$ is the average of
their electric charges, and $F_2^{\NS}$ is the corresponding
combination of photon-induced DIS structure functions.
We then consider a so--called physical factorization scheme~\cite{Diemoz:1987xu,Catani:1995ze,Catani:1996sc,Altarelli:1998gn,Lappi:2023lmi},
whose prototype is the DIS scheme of Ref.~\cite{Diemoz:1987xu}. This is defined by imposing the
 condition
\begin{equation}\label{eq:disfact1}
   C^{\rm NS,\,DIS}\left(\alpha_s(Q^2),
   \frac{Q^2}{\mu_f^2}\right)\Bigg|_{\mu^2_f=Q^2}=1
   \end{equation}
to all perturbative orders~\footnote{This is analogous to a
renormalization condition in an on-shell
scheme, such as that commonly adopted to renormalize
the electric charge in QED, in which it is required that the full
vertex function $\Gamma^\mu(q^2)$ at zero momentum transfer coincides to
all perturbative orders with its leading-order
expression: $\Gamma^\mu(0)=\gamma^\mu$.}.

The DIS and \msbar{} schemes are related by a finite change of scheme 
\begin{equation}\label{eq:finren}
   C^{\rm NS,\,DIS}\left(\alpha_s(Q^2), \frac{Q^2}{\mu_f^2}\right)\Bigg|_{\mu^2_f=Q^2}= Z^{\rm
     NS,\,DIS}(\alpha_s(Q^2))  \otimes C^{\rm
     NS,\,\mmsbar}\left(\alpha_s(Q^2),\frac{Q^2}{\mu_f^2}\right)\Bigg|_{\mu^2_f=Q^2},
   \end{equation}
and the renormalization condition, Eq.~(\ref{eq:disfact1}), immediately
implies that
\begin{align}
  \label{eq:distomsbara}
Z^{\rm
     NS,\,DIS}(\alpha_s(Q^2)) =
\left[C^{\rm NS,\,\mmsbar}\right]^{-1}\left(\alpha_s(Q^2)\right),
\end{align}
where  $\left[C^{\rm
    NS,\,\mmsbar}\right]^{-1}$ denotes the functional inverse of the
coefficient function upon convolution, i.e., the distribution (in
general) such that
\begin{equation}\label{eq:functinv}
\left[C^{\rm NS,\,\mmsbar}\otimes\left[C^{\rm
      NS,\,\mmsbar}\right]^{-1}\right](x)=\delta(1-x)\,.
\end{equation}

Substituting the expression of the DIS-scheme coefficient function,
Eq.~(\ref{eq:disfact1}), in the leading-twist factorization,
Eq.~(\ref{eq:disfact0}), we immediately get
\begin{align}\label{eq:disfactnsa}
&  \frac{1}{x} F_2^{\NS}(x,Q^2)= \frac{1}{x} F_2^{\rm\NS,\,
    LT}(x,Q^2)+ O\left(\frac{\Lambda^2}{Q^2}\right),  \\\label{eq:disfactnsb}
&\quad \frac{1}{x} F_2^{\rm \NS,\, LT}(x,Q^2)=  \langle e_i^2\rangle
 {f^{\NS,\,\rm DIS}}(x,Q^2) .
\end{align}
This means that the DIS-scheme PDF provides, up to the constant
prefactor $\langle e_i^2\rangle$, the leading twist
expression of the structure function, hence its positivity properties
are determined by those of the leading twist expression of the
structure function,   $F_2^{\rm \NS,\,
    LT}(x,Q^2)$.
Furthermore, the relation between the DIS and \msbar{} schemes,
Eq.~(\ref{eq:finren}), implies that the 
respective PDFs are related by
\begin{align}
  \label{eq:distomsbarb}
f^{\NS,\,\mmsbar}(x,Q^2)&=
\left[C^{\rm NS,\,\mmsbar}\right]^{-1}\left(\alpha_s(Q^2)\right) \otimes   {f^{\NS,\,\rm DIS}}(x,Q^2) \,.
\end{align}
Hence, the  positivity properties of the  \msbar{} PDF are in turn
determined by those of the
structure function 
and the inverse of the perturbatively
computable coefficient function $C^{\rm NS,\,\mmsbar}$ without any
reference to non-perturbative information.

Now, of course the nonsinglet structure function is not positive,
however, one can choose a set of physical observables that fully
determines the  complete set of PDFs, and specifically
choose~\cite{Altarelli:1998gn,Candido:2020yat} a set of physical observables that are
linear in the PDF, such as deep-inelastic structure functions which at
leading order define the quark and a Higgs production process which
defines the gluon.
Equation~\eqref{eq:disfact0} then generalizes to
\begin{align}\label{eq:disfact}
\sigma (x,Q^2)= \sigma_0  C^{\rm \mmsbar}\left(\alpha_s(Q^2), \frac{Q^2}{\mu_f^2}\right) \otimes 
 f^{\mmsbar}(\mu_f^2) +O\left(\frac{\Lambda^2}{Q^2}\right),
 \end{align}
where $\sigma (x,Q^2)$ is now a vector of physical observables,
$f^{\mmsbar}(x,Q^2)$ is a vector of PDFs, $ C^{\rm \mmsbar}$ is a
matrix of coefficient functions and  $\sigma_0$ is a diagonal
matrix of coefficients (analogous to the coefficient $\langle
e_i^2\rangle$ in Eq.~(\ref{eq:disfact})) chosen in such a way that to
leading order 
$C^{\rm \mmsbar}=\mathbb{I}$, the identity matrix.

Equations~(\ref{eq:disfactnsa}-\ref{eq:disfactnsb}) then become
\begin{align}\label{eq:physfacta}
&  \sigma (x,Q^2)=\sigma^{\rm LT} (x,Q^2)+ O\left(\frac{\Lambda^2}{Q^2}\right)  \\ \label{eq:physfactb}
&\quad  \sigma^{\rm LT} (x,Q^2)=\sigma_0  f^{\rm PHYS}(x,Q^2),
\end{align}
and the  \msbar{} PDFs are then given by the matrix
generalization of Eq.~(\ref{eq:distomsbarb}), namely
\begin{align}\label{eq:distomsbar}
  f_i^{\mmsbar}(x,Q^2)
=\left[C^{\mmsbar}\right]_{ij}^{-1}\left(\alpha_s(Q^2)\right) \otimes f_j^{\rm PHYS}(Q^2) \,.
\end{align}
But, because of Eq.~(\ref{eq:physfactb}), $f_j^{\rm PHYS}(Q^2)$ is now
a set of physical observables computed at leading twist, so, unless
the leading-twist approximation breaks down, the
physical scheme PDFs are positive, and whether the  \msbar{} PDFs are
also positive, and where, can be determined by studying the
properties of the  matrix of
perturbative coefficients $\left[C^{\mmsbar}\right]_{ij}$ and its inverse.

Note that Eq.~(\ref{eq:distomsbar}) is true regardless of the accuracy of the
leading-twist approximation. The PDFs are defined as the matrix
elements of leading-twist operators~\cite{Collins:1981uw} (see
e.g.\ Eq.~(2.2) of Ref.~\cite{Candido:2020yat}), and  indeed, their moments
are matrix elements of leading-twist operators. 
Equation~(\ref{eq:physfactb}) then means that the matrix element of
these leading twist operators, in the physical scheme, provides the
result of the leading-twist computation of a set of physical
observables.
Equation~(\ref{eq:distomsbar}) relates these matrix elements to those of the
same operators, but now in the \msbar{} scheme. Whether the leading
twist expression Eq.~(\ref{eq:physfacta}) of the full set of
cross-sections $\sigma (x,Q^2)$ is accurate, i.e.\ whether it
does or does not provide a good approximation to an exact computation
(that would also include higher twist corrections),
is a separate issue. We will discuss this issue, which clearly has
phenomenological implications, in the last section.

The question that we wish to address now is whether, within
the realm of validity of the leading twist approximation, in which the
physical scheme PDF is surely positive, the positivity is preserved
upon transformation to the \msbar{} scheme. This is tantamount to
asking whether the positivity 
of $f^{\rm PHYS}(Q^2)$ is preserved
upon convolution with $\left[C^{\mmsbar}\right]^{-1}\left(\alpha_s(Q^2)\right)$.
In the perturbative domain we have 
\begin{align}\label{eq:msbartodisa}
f^{\rm PHYS}(x,Q^2)=\left[1
 +\frac{\alpha_s}{2\pi}  
    C^{(1),\,\mmsbar}\otimes\right]{f^{\mmsbar}}(Q^2)+\mathcal{O}(\alpha_s^2) \,,  
\end{align}
and the perturbative inverse is just 
\begin{align}\label{eq:distomsbarc}
f^{\mmsbar}(x,Q^2)=\left[1-\frac{\alpha_s}{2\pi} 
    C^{(1),\,\mmsbar}\otimes\right]f^{\rm PHYS}(Q^2)+\mathcal{O}(\alpha_s^2) \,.
\end{align}
Hence the positivity condition is
\begin{equation}\label{eq:noposcond}
\left|\frac{\alpha_s}{2\pi}     C^{(1),\mmsbar}\otimes f^{\rm PHYS}\right| \leq
  |f^{\rm PHYS}|,
\end{equation}
  which should be
understood as a set of conditions for each component of the vector of
PDFs $f^{\rm PHYS}$. 

The perturbativity condition, in turn, which ensures that the
manipulation from Eq.~(\ref{eq:msbartodisa}) to
Eq.~(\ref{eq:distomsbarc}) is justified, has the same form as
Eq.~(\ref{eq:noposcond}), but with $f^{\rm PHYS}$ replaced by
$f^{\mmsbar}$. It was already noted in Ref.~\cite{Candido:2020yat}
that these conditions are manifestly violated as $x\to1$, because the
$\mathcal{O}(\alpha_s)$ term on the r.h.s.\ of
Eq.~(\ref{eq:msbartodisa}) contains contributions that are not
uniformly small for all $x$, and are in fact unbounded as $x\to1$.
We must therefore deal with these contributions before we can discuss
the perturbative positivity condition Eq.~(\ref{eq:noposcond}).

The issue was already addressed in
Ref.~\cite{Candido:2020yat}, by
performing an exact inversion of these terms in the $x\to1$ region.
We now revisit this argument,
in particular, by explicitly discussing the treatment of
distributional contributions.
To this purpose, we write the coefficient function up to
$\mathcal{O}(\alpha_s)$ by first, separating off, as in 
Ref.~\cite{Candido:2020yat}, the contribution proportional to a Dirac
$\delta$ so that the remainder only contains ordinary functions and
$+$ distributions, and then further splitting this remainder in a
``finite'' and a ``divergent'' contribution: 
\begin{align}\label{eq:decompa}
C^{\mmsbar}(x)&= \delta(1-x) +\frac{\alpha_s}{2\pi}\left[ \delta(1-x)
  \Delta^{(1)}+C_F^{(1),\,\mmsbar}(x)+C_D^{(1),\,\mmsbar}(x)\right] + \mathcal{O}(\alpha_s^2)\\\label{eq:decompb}
&=\left[\delta(1-x) +\frac{\alpha_s}{2\pi}\left[ \delta(1-x)
  \Delta^{(1)}+C_F^{(1),\,\mmsbar}(x)\right]\right]\nonumber\\&\qquad\otimes
\left[\delta(1-x) +\frac{\alpha_s}{2\pi} C_D^{(1),\,\mmsbar}(x)\right] + \mathcal{O}(\alpha_s^2).
\end{align}
In Eq.~\eqref{eq:decompa} $ \Delta^{(1)}$ is a diagonal
matrix of constants whose explicit expressions  are given in
Ref~\cite{Candido:2020yat}; $C_F^{(1),\,\mmsbar}(x)$ are contributions such that
$\lim_{x\to1}R_F(x)$ is finite, where $R_F(x)$ is defined as
\begin{equation}\label{eq:rf}
  R_F(x)\equiv\frac{ [C_F^{(1),\,\mmsbar}\otimes f](x)}{f(x)},
\end{equation} 
while $C_D^{(1),\,\mmsbar}(x)$ are contributions such that, defining
\begin{equation}\label{eq:rd}
  R_D(x)\equiv\frac{ [C_D^{(1),\,\mmsbar}\otimes f](x)}{f(x)},
\end{equation}
then $\lim_{x\to1}R_D(x)=\infty$. Contributions 
$C_D^{(1),\,\mmsbar}(x)$ are henceforth referred to as divergent contributions.

As well known,
divergent contributions are present to all perturbative orders in the \msbar{} scheme in the
diagonal elements of the matrix of coefficient functions: they are proportional to
$S_k(x)=\left[\frac{\ln^k(1-x)}{1-x}\right]_+$ 
and they are due to soft
gluon emission~\cite{Sterman:1986aj,Catani:1989ne}. Note that in
principle unbounded contributions are also present when $x\to0$. These
could be handled in the same way, but we do not discuss them since
PDFs grow large and positive as $x\to0$ due to high-energy
logs~\cite{Altarelli:2008aj}, so positivity constraints are of no importance. 

Clearly, there is some latitude in defining the separation of
Eq.~\eqref{eq:decompa} of the coefficient function into an $C_F^{(1),\,\mmsbar}$ term
and a $C_D^{(1),\,\mmsbar}$ term, since we can always subtract a finite contribution from
$C_F^{(1),\,\mmsbar}$ and include it into $C_D^{(1),\,\mmsbar}$: for example,
$\left[\frac{1+x^2}{1-x}\right]_+=\frac{1+x}{(1-x)_+}+\frac{3}{2}\delta(1-x)$.
We define  $C_D^{(k),\,\mmsbar}(x)$ as the
contribution that includes all and only the $S_k(x)$ terms.
With this definition, up to $\mathcal{O}(\alpha_s)$ we have
\begin{equation}\label{eq:lld}
  \left[C_D^{(1),\,\mmsbar}\right]_{ij}(x)= c_i\delta_{ij}\left(
  2\left[\frac{\ln(1-x)}{1-x}\right]_+-\frac{3}{2} 
  \left[\frac{1}{1-x}\right]_+\right)=c_i\delta_{ij} \left[\frac{\ln(1-x)}{1-x}\right]_++
\hbox{NLL}(1-x),
\end{equation}
where $i=q,g$, $c_q=C_A$ and $c_g=C_F$ (see e.g.~\cite{Candido:2020yat}).

Having written the coefficient function as in Eq.~\eqref{eq:decompb} its inverse
can be determined by performing the inversion perturbatively for all
terms but $C_D^{(1),\,\mmsbar}(x)$: namely
\begin{align}\label{eq:cdecomp}
\left[C^{\mmsbar}\right]^{-1}=&
\left[\delta(1-x) -\frac{\alpha_s}{2\pi}\left(  \delta(1-x)
  \Delta^{(1)}+C_F^{(1),\,\mmsbar}(x)\right)\right]\nonumber\\&\qquad \otimes
\left[\delta(1-x) +\frac{\alpha_s}{2\pi}
  C_D^{(1),\,\mmsbar}(x)\right]^{-1} + \mathcal{O}(\alpha_s^2),
\end{align}
where now  $C_D^{(1),\,\mmsbar}(x)$ must be computed by requiring it
to satisfy Eq.~\eqref{eq:functinv} exactly, i.e., not just up to higher
order perturbative corrections. We refer to this as the
exact inverse: note that this is of course only exact within the
limitations of leading-twist factorization.

The exact inverse of the divergent contribution was computed already
in Ref.~\cite{Candido:2020yat} to leading logarithmic accuracy, and it
was shown to be finite in the $x\to 1$ limit. This was argued to be
in agreement with the
expectation that the inverse of a divergent coefficient is actually
finite, based on Mellin-space considerations. Namely, the Mellin transform of a
divergent coefficient function diverges as the Mellin variable
$N\to\infty$, but the Mellin-space inverse is just the reciprocal, so
the Mellin transform of the inverse vanishes as $N\to\infty$, implying that
the inverse is finite.

We wish now to make this argument 
more explicit by fully working out distributional
contributions and explicitly proving their positivity.
To leading logarithmic accuracy we have~\cite{Candido:2020yat}
\begin{align}
  \label{eq:lldinv}
  &\left[\delta(1-x) +\frac{\alpha_s}{2\pi}
  C_D^{(1),\,\mmsbar}(x)\right]^{-1}_{ij}=\delta_{ij}\delta\left(1-x\right) \nonumber \\
  &\qquad
  -2\delta_{ij}c_i\frac{\alpha_s}{2\pi} \left(
  \frac{\ln(1-x)}{\left(1+c_i\frac{\alpha_s}{2\pi}\ln^2(1-x)\right)^2}  \frac{1}{1-x}\right)_++
  \hbox{NLL}(1-x).
\end{align}
Note that the term in brackets has an integrable singularity as
$x\to1$, but it is nevertheless expressed as a $+$ distribution
because it is obtained by resummation of a series of contributions
each of which has a $+$ distribution taming a non-integrable
singularity.  

We can now discuss the positivity of the various contributions by
working out their action on a test PDF $f(x)$. Note that when substituting the inverse in
Eq.~\eqref{eq:distomsbara} the PDF that is acted upon is 
guaranteed to be positive, because it is a physical observable. We use
the identity 
\begin{align}
  \label{eq:convplus}
  \left[f_+\otimes g\right]\left(x\right)&=\int_x^1 \frac{dz}{z} f\left(z\right)_+ g\left(\frac{x}{z}\right) \nonumber \\
   &= \int_x^1 \frac{dz}{z}f\left(z\right)g\left(\frac{x}{z}\right) - g\left(x\right)\int_0^1 dz \, f\left(z\right).
\end{align}
We get
\begin{align}\label{eq:finreninv}
  \bar f_i(x)&\equiv \left[Z_D\otimes f\right]_i\left(x\right)\\  
  &=\sum_j \left[\delta(1-x) +\frac{\alpha_s}{2\pi}
  C_D^{\left(1\right),\,\mmsbar}\left(x\right)\right]^{-1}_{ij}
  \otimes f_j\left(x\right)\nonumber \\
  &=
  2c_i\frac{\alpha_s}{2\pi} \int_x^1\frac{dz}{z} 
  \frac{\ln\left(\frac{1}{1-z}\right)}{\left(1+c_i\frac{\alpha_s}{2\pi}
  \ln^2\left(1-z\right)\right)^2} 
  \frac{f_i \left(\frac{x}{z}\right)}{1-z}\nonumber \\
  &\qquad 
  + f_i\left(x\right)\left[1+2c_i\frac{\alpha_s}{2\pi}
  \int_0^1 \frac{dz}{1-z}\frac{\ln\left(1-z\right)}
  {\left(1+c_i\frac{\alpha_s}{2\pi}\ln^2\left(1-z\right)\right)^2}\right]\nonumber
  \\ \label{eq:expinvb}
  &=
  2c_i\frac{\alpha_s}{2\pi} \int_x^1\frac{dz}{z} 
  \frac{\ln\left(\frac{1}{1-z}\right)}{\left(1+c_i\frac{\alpha_s}{2\pi}\ln^2(1-z)\right)^2} 
  \frac{f_i \left(\frac{x}{z}\right)}{1-z}.
\end{align}
In Eq.~(\ref{eq:finreninv}) we have viewed the  exact inverse of the
divergent coefficient function as a finite renormalization $Z_D$ that
transforms the PDF $f$ into a new PDF $\bar f$.
It is clear that, as argued in Ref.~\cite{Candido:2020yat},
Eq.~\eqref{eq:expinvb} vanishes
as $x\to1$, and in fact it behaves as $\frac{1}{\ln^2 (1-x)}$ as
$x\to1$, as one might expect given that the convolution of the
\msbar{}  coefficient function with the PDF has double log $\ln^2(1-x)$ behavior. 
Furthermore,  Eq.~\eqref{eq:expinvb} is manifestly positive given that
the integrand
is the product of the positive physical scheme PDF times a
positive function.

We can now get back to the positivity argument. We substitute the
expression for the inverse of the coefficient function in which the
divergent terms have been inverted exactly,
Eq.~(\ref{eq:cdecomp}), in the expression of
the \msbar{} PDFs,  Eq.~(\ref{eq:distomsbar}). We then observe that $f_j^{\rm PHYS}(Q^2,x)$ are positive,
monotonically decreasing functions of $x$ for all scales in the
perturbative domain - e.g.\ structure functions are monotonic in $x$
for $Q^2\gtrsim1$~GeV$^2$. It then follows that transforming $f_j^{\rm
  PHYS}(Q^2,x)$ through the finite renormalization factor
$Z_D$ implicitly defined in
Eq.~(\ref{eq:finreninv}) preserves its positivity, and thus the
condition for positivity of the \msbar{} PDF is given by
Eq.~(\ref{eq:noposcond}), with now  $f_j^{\rm
  PHYS}$ replaced by $\bar f_j^{\rm
  PHYS}$.

We can make the condition independent of the PDF by a series of
majorizations. First, we note that, because the PDF is monotonically decreasing,
\begin{align}\label{eq:posconda}
\left|  \sum_j \int_x^1\frac{dy}{y} C_{F,ij}^{(1),\mmsbar}(y)
 f_j^{\rm PHYS}\left(\frac{x}{y}\right)\right| &\le     \sum_j
 f_j^{\rm PHYS}(x) \int_x^1\frac{dy}{y} \left| C_{F,ij}^{(1),\mmsbar}(y)\right|\\\label{eq:poscondaa}
 &\le f^{\rm PHYS,\,{\rm MAX}}(x)  \sum_j  \int_x^1\frac{dy}{y} \left|
 C^{(1),\mmsbar}_{F,ij}(y)\right|
\end{align} 
where we
have denoted
by $f^{\rm PHYS,\,{\rm MAX}}(x)$  the largest component  of the vector
$f^{\rm PHYS}(x)$.
Using the majorization Eq.~\eqref{eq:posconda} in the perturbativity
condition Eq.~\eqref{eq:noposcond}, using the decomposition
Eq.~(\ref{eq:decompb}), and neglecting the $\Delta^{(1)}$
contribution, which is a finite renormalization of $f^{\rm PHYS}$, we end up with the  condition 
\begin{equation}\label{eq:poscondfin}
  \frac{\alpha_s}{2\pi}
    \sum_j  \overline{{\cal C}}_{ij}(x)  \ll
      \frac{f^{\rm PHYS}(x)_i}{f^{\rm PHYS,\,{\rm
            MAX}}(x)}\le 1,
\end{equation}
where we have defined cumulants
\begin{equation}\label{eq:cumulant}
\overline{{\cal C}}_{ij}(x)  = \int_x^1\frac{dy}{y}
 \left| C^{(1),\mmsbar}_{F,ij}(y)\right|.
 \end{equation}
This is indeed independent of the PDF.

\begin{figure}[t]
  \begin{center}
    \includegraphics[width=0.45\linewidth]{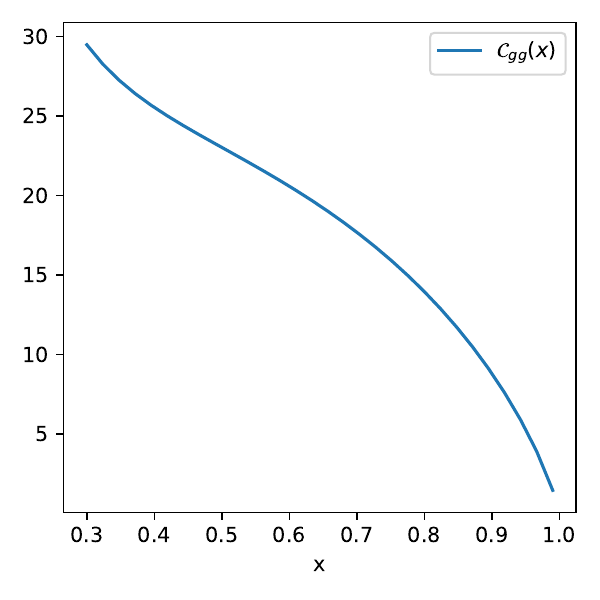}
    \includegraphics[width=0.45\linewidth]{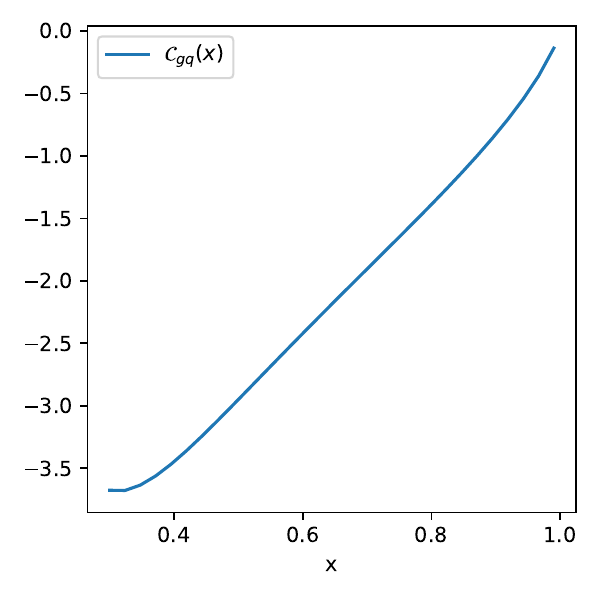}\\
    \includegraphics[width=0.45\linewidth]{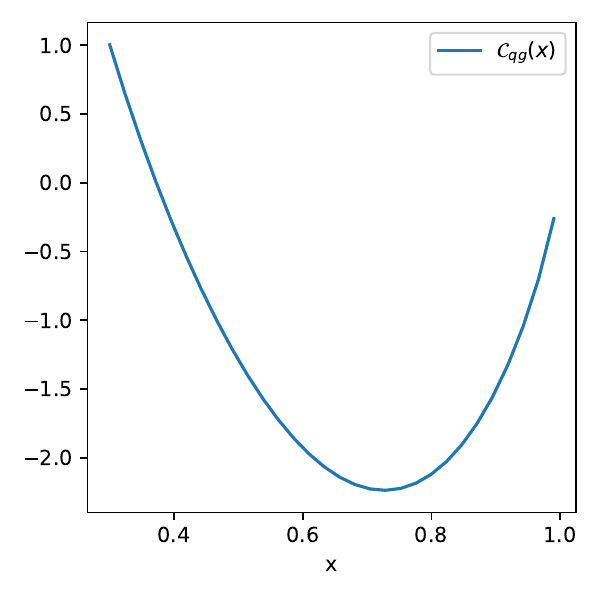}
    \includegraphics[width=0.45\linewidth]{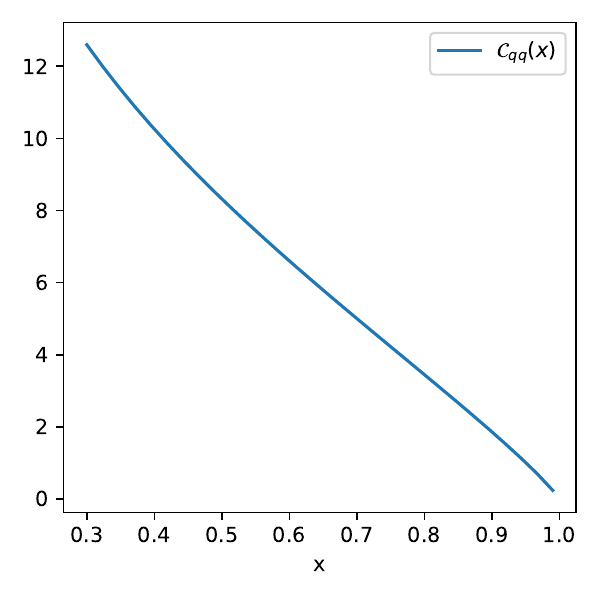}
    \caption{\small The NLO integrated coefficient functions  ${\cal C}_{ij}(x)$ Eq.~\eqref{eq:intcumulant}. The
      gluon-gluon, gluon-quark, quark-gluon and quark-quark entries
      are shown from left to right and from top to bottom.  
    \label{fig:pert} }
  \end{center}
\end{figure}
At NLO we only need two independent physical processes in order to fix
the factorization scheme~\cite{Altarelli:1998gn}, and we then have a
two-by-two quark-gluon matrix of
coefficient functions  $C^{(1),\mmsbar}_{ij}$ and thus of finite terms
$C^{(1),\mmsbar}_{F,ij}$ and cumulants $\overline{\cal C}_{ij}$. We choose the same processes as in
Ref.~\cite{Altarelli:1998gn} in order to define the physical scheme,
namely deep-inelastic scattering and Higgs production in gluon
fusion. The integrated coefficient functions
\begin{equation}\label{eq:intcumulant}
 {\cal C}_{ij}(x)  = \int_x^1\frac{dy}{y}
  C^{(1),\mmsbar}_{F,ij}(y)
 \end{equation}
computed with this choice are displayed in
Fig.~\ref{fig:pert}. It
is easy to check that for these processes the finite renormalization
from the $\Delta^{(1)}$ contribution Eq.~(\ref{eq:decompb}) is always
positive for the gluon, and for the quark positive for $Q\gtrsim 1$~GeV.

We can now discuss the positivity region quantitatively. First, we
note that the region in which monotonic decrease of physical
observables sets in can be very conservatively estimated to be
$Q^2\gtrsim5$~GeV$^2$. Indeed, as already observed, deep-inelastic
structure functions display monotonic behavior with a characteristic
small $x$ rise for $Q^2\gtrsim1$~GeV$^2$~\cite{H1:2015ubc}. In fact,
for $Q^2\gtrsim5$~GeV$^2$ \msbar{}  quark and gluon PDFs are
monotonic, with a small $x$ rise which is driven by leading-order
perturbative evolution, and thus essentially
scheme-independent. Indeed,  already for   $Q^2\gtrsim10$~GeV$^2$
for all $x\lesssim0.1$~\cite{Caola:2008xr} PDFs satisfy to good accuracy the
double-scaling behavior~\cite{Ball:1994du} which characterizes
leading-order small-$x$ perturbative evolution. This behavior of the
PDFs also implies that for $Q^2\gtrsim5$~GeV$^2$ positivity issues
only possibly arise at large $x$.

We consequently focus on the behavior of the cumulants
 and the condition Eq.~(\ref{eq:poscondfin}) at
large $x$. It is clear that as $x$ decreases, because of the rapid
decrease of $f_i(x)$ as $x$ increases, the majorization
Eq.~\eqref{eq:posconda} becomes increasingly more conservative,
leading to an increasingly restrictive condition. A realistic
condition must thus be imposed at sufficiently large $x$.  
Figure~\ref{fig:pert} shows that, in  practice, the condition is dominated by the
gluon-gluon entry  ${\cal C}_{gg}(x)= \overline{{\cal C}}_{gg}(x)$ .
Imposing  very conservatively the condition at the valence peak
$x\sim 0.3$ and
requiring  $\frac{\alpha_s}{2\pi}  {\cal C}_{gg}(x)<1$, we get
$\alpha_s(Q^2)\lesssim0.2$,  corresponding again to 
$Q^2\gtrsim5$~GeV$^2$.  Imposing the condition at a higher $x$ value would lead to a
lower scale, because of the rapid decrease of the cumulant, but then
at some point
close to threshold the observables used to define the PDFs become
non-perturbative and partly ill-defined: for example,  for
$x\gtrsim0.9$  structure functions are no longer given by the
continuum and enter the resonance region. Indeed,
in the NNPDF4.0 PDF determination
positivity conditions are imposed at  $Q^2=5$~GeV$^2$. Clearly, this
should be taken as a very  conservative semi-quantitative
estimate.

\section{The positivity domain}
\label{sec:conc}

We can now collect the results of the previous two sections, and
discuss their implications for PDF positivity. In Section~\ref{sec:fact}
we have shown that, in agreement with
Ref.~\cite{Collins:2021vke},
PDFs always turn positive
at high enough scale thanks to logarithmically enhanced contributions
of perturbative origin, and, correspondingly, turn negative at a low enough scale, as pointed out
in Ref.~\cite{Collins:2021vke}. However, the value of this scale is set by
non-perturbative physics and thus difficult to determine precisely.

In Sect.~\ref{sec:pospos} we have then shown  that by considering the relation
between the  \msbar{} factorization scheme and a physical
factorization scheme, we can show that positivity of the
physical-scheme PDFs is inherited by the \msbar{} PDFs, for a scale
that is large enough that perturbativity holds. The value of this
scale can be
estimated on the basis of purely perturbative considerations, in terms
of the properties of the perturbative expansion of a set of \msbar{}
coefficient functions used to define the physical scheme.

It is important to understand what is nontrivial in this conclusion,
and what are its implication. The nontrivial fact is that within the
perturbative domain the \msbar{} PDFs inherit the positivity of
physical-scheme PDFs. For instance, if the coefficient of the
$O(\alpha_s)$ term on the r.h.s.\ of Eq.~(\ref{eq:lldinv}) had the
opposite sign, this would not be the case, and the  \msbar{} PDFs
would always turn negative for large enough $x$ in the perturbative
domain: hence the scale at which the PDF turns positive in such case
would be $x$-dependent, and in fact it would become arbitrarily large
as $x\to1$. The physical origin of this sign was discussed in
Ref.~\cite{Candido:2020yat}: it is due to the fact that   the
\msbar{} scheme amounts to an over-subtraction of collinear
singularities, but these in the diagonal channel have a negative sign,
corresponding to the well-known Sudakov suppression of QCD processes
in the soft limit~\cite{Amati:1980ch}.

The conclusion that the \msbar{} PDFs inherit the positivity of
physical-scheme PDFs has nontrivial implications for PDF
phenomenology. Assume that in a PDF determination, by using
Eq.~(\ref{eq:refacb}), it is found that some of the fitted \msbar{} PDFs
$f^{\rm fit}_j$ must turn 
negative, for some value of $x$, in 
order to guarantee agreement with data. Note that this value of $x$ could be
outside the region of $x$ values that are probed at leading order: for
example, it could be some large $x$ value that contributes upon
convolution to a structure function which instead is measured for
smaller values of the Bjorken variable. In that case, the structure
function could well remain positive even with some of the \msbar{}
PDFs turning negative, say
at large $x$. Now, the implication of the argument
presented here is that, in such a case, these fitted $f^{\rm fit}_j$
cannot be identified with the true \msbar{} PDFs $f_j$.
Indeed, our result, that if the
physical scheme PDFs are positive, then the  \msbar{} PDF also are,
implies that if the \msbar{} PDFs are {\it not} positive, then the physical
scheme PDF also are not positive. But the physical scheme PDFs coincide
with positive physical observables up to higher twist terms, hence if
they are negative then this means that higher-twist contributions must
be present.

However, the PDF is a leading--twist quantity, while
higher-twist contributions correspond to multiparton
correlations. Hence, the measured $f^{\rm fit}_j$ that have turned negative
differ from the true leading-twist
PDFs $f_j$, and do not enjoy their properties. For instance, the
second moment of $f^{\rm fit}_j$ will not provide the momentum fraction carried by
the $j$-th  parton, because the momentum fraction is the matrix
element of the twist-two energy momentum tensor. Or, to give another
example, $f^{\rm fit}_j$ are not universal, and thus  cannot  be used
to predict, say, 
production of a heavy BSM gauge bosons, which would probe the large
$x$ dependence of $f^{\rm fit}_j$.
Our results therefore provide an effective criterion to detect the
presence of higher--twist corrections, even in regions that are
inaccessible by leading-order kinematics.

Our results are wholly within the domain of a leading-twist
perturbative approach.
A more fundamental approach could possibly start from an assessment of the non-perturbative operator matrix
element in terms of which the PDF is defined. If this were known,
perturbative computations along the lines of Sect.~\ref{sec:fact}
would enable a determination of the scale above which positivity
holds. Such a non-perturbative computation would be a priori much more
powerful, and might even lead to insights on the region of validity of
the leading-twist approximation and of factorization based upon
it. However, this is beyond what can be achieved at present, and well
beyond the scope of this investigation.

A further limitation of this paper is that all of its results
apply to massless quarks. It
would be interesting to extend the discussion to the case of heavy
quarks. This would be particularly interesting in view of recent
results providing evidence for an intrinsic charm component of the
proton~\cite{Ball:2022qks}, and also in view of  subtle issues related
to the validity of factorization for hadronic processes with heavy
quarks in the initial state~\cite{Caola:2020xup}. This will be left to
future investigations.

{\bf Acknowledgments:} We thank Luigi Del Debbio, Ted Rogers and Nobuo Sato
for discussions. We are especially grateful to  Aleksander Kusina,
Andrzej Si\'odmok and in particular James Whitehead for critical input
and for pointing out errors in the discussion of
Eq.~(\ref{eq:expinvb}), in
Eqs.(\ref{eq:poscondaa}-\ref{eq:poscondfin}) and in Fig.~\ref{fig:pert} in the original version of this paper.
T.~G.\ is supported by NWO via an ENW-KLEIN-2 project.
F.~H.\ is supported by the Academy of Finland
project 358090 and is funded as a part of the Center
of Excellence in Quark Matter of the Academy of Finland, project 346326.

\appendix
\section{Perturbative computation of the bare PDF}
\label{sec:app1}

The explicit expression of the fully bare quark PDF of an off-shell
massless quark can be obtained by defining the PDF as the matrix
element of a Wilson line operator~\cite{Collins:1981uw}, see Eq.~(2.2)
of Ref.~\cite{Candido:2020yat}, and evaluating this matrix element in a free off-shell massless
quark state. The matrix element can be computed perturbatively using the Feynman rules for
Wilson lines (see Sect.~7.6 of Ref.~\cite{Collins:2011zzd}). The
computation is then similar to that presented in Sect.~9.4.3 of this
reference, in the case of a massless, off-shell quark.
Working in dimensional regularization we get, for a target quark with
virtuality $M^2$, 
\begin{align}
  \label{eq:bare}
  f^{M,\,0}_{q/q}&\left(x,\mu^2, \epsilon\right) = \delta(1-x)\,Z \nonumber \\
  & + 2\pi\,g^2\mu^{2\epsilon}C_F 
  \left(1-x\right)
  \int\frac{d^{2-2\epsilon}\bf{k}_\text{T}}{\left(2\pi\right)^{4-2\epsilon}} \, 
  \frac{\left(1-\epsilon\right)\left(k_\text{T}^2 + x^2M^2 \right)}
  {\left(k_\text{T}^2 - x\left(1-x\right)M^2\right)^2} \nonumber \\
  & +  4\pi\,g^2\mu^{2\epsilon}C_F \biggl[\frac{x}{\left(1-x\right)}
  \int \frac{d^{2-2\epsilon}\bf{k}_\text{T}}{\left(2\pi\right)^{4-2\epsilon}}
  \frac{1}{k_\text{T}^2 - x\left(1-x\right)M^2} \nonumber \\
  & \qquad\qquad -  \,\delta\left(1-x\right)
  \int_0^1 d\alpha\frac{\alpha}{\left(1-\alpha\right)}
  \int \frac{d^{2-2\epsilon}\bf{k}_\text{T}}{\left(2\pi\right)^{4-2\epsilon}}
  \frac{1}{k_\text{T}^2 -\alpha\left(1-\alpha\right)M^2}\biggr]
\end{align}
where  $Z$ is the residue in the pole of the quark propagator, given by
\begin{align}
  \label{eq:residue}
  &Z = 1  - g^2\mu^{2\epsilon} C_F 
  \left(2-2\epsilon\right)  
  \int_0^1 dx\, \left(1-x\right) \int  
  \frac{d^{4-2\epsilon}{\bf k}}{\left(2\pi\right)^{4-2\epsilon}} \frac{1}
  {\left(-k^2 -x\left(1-x\right)M^2\right)^2} .
\end{align}
A direct computation of the different terms gives
\begin{align}
  f^{M,\,0}_{q/q}\left(x,\mu^2, \epsilon\right) = &
  \delta\left(1-x\right) 
  +\frac{\alpha_s}{2\pi}C_F\left\{ 
    \Gamma\left(\epsilon\right) \,
    \left[\left(\frac{\Delta^2\left(x\right)}{4\pi\mu^2}\right)^{-\epsilon}
      p_{qq}\left(x\right)\right]_+ +  f_f(x)\right\},
\end{align}
where $ p_{qq}\left(x\right)$ is as in Eq.~\eqref{eq:bare1},
$\Delta^2\left(x\right)$ is given by Eq.~\eqref{eq:mass_param},
$\alpha_s = \frac{g^2}{4\pi}$,  and  $f_f(x)$ is a finite contribution
given by
\begin{equation}\label{eq:finite}
  f_f(x)=
    \frac{1}{2}\delta\left(1-x\right) +
    x-2.
\end{equation}

\bibliographystyle{UTPstyle}
\bibliography{pos2}

\end{document}